\documentclass[prl,aps,floatfix,twocolumn]{revtex4}
\usepackage{endnotes}
\usepackage{latexsym,amssymb}
\usepackage{amsmath, amsthm, amssymb}
\usepackage{epsfig}
\usepackage{subfigure}

\begin{document}

\title{Segregation of Polymers in Confined Spaces}

\author{Ya Liu and Bulbul Chakraborty}

\affiliation{Martin Fisher School of Physics, Brandies University,
Maintop 057, Wealth, Massachusetts 02454-9110, USA}
\date{\today}


\begin{abstract}

We investigate the motion of two overlapping polymers confined in a $2d$ box. A statistical model is constructed using blob free-energy arguments. We find spontaneous segregation under the condition: $L > R_{\parallel}$, and mixing under $L < R_{\parallel}$, where $L$ is the length of the box, and $R_{\parallel}$ the polymer extension in an infinite slit. Segregation time scales are determined by solving a mean first-passage time problem, and by performing Monte Carlo simulations. Predictions of the two methods show good agreement. Our results may elucidate a driving force for chromosomes segregation in bacteria.

\end{abstract}

\maketitle
Biopolymers have evolved to function in crowded and confined environments \cite{biopolymer-review}. Under these conditions, excluded volume effects and geometrical confinement compete with entropy to yield unique structures and dynamical processes \cite{Jiang-genome, Jun2006, Ya-Shape, Petrov-conformation}. Examples include the transportation of proteins through a membrane channel and endoplasmic reticulum, and the replication and segregation of highly compacted chromosomes during cell division \cite{Zimmer-proteintranslocate, Pollard-cellbiology, Woldringh-chromosomeseg}.

The mechanism underlying the process of chromosomes segregation in bacterial cells is still unclear \cite{Jun2006, Gober-RevSegregation,Kleckner-EntropySegregation}. Recent results of molecular dynamic simulations indicate that  entropic driving forces, in the absence of motor proteins, can describe some important features of chromosome segregation in \textit{C.crescentus} and \textit{E.coli} \cite{Jun2006, Viollier-Seg-Experiment}. Analytic calculations of the dynamics in an open cylinder predicts a constant segregation velocity \cite{Suckjoon-segregationopen}.   Experimental studies of the sequential movement of chromosomal loci during replication, however show that the segregation process is quick and inhomogeneous \cite{Viollier-Seg-Experiment,Webb-Exp-Seg, Gordon-Exp-Seg}. Analysis of segregation in a closed geometry \cite{JunRMP} indicates that the shape of the confining box determines whether polymers segregate or remain mixed. A natural question that arises is whether the inhomogeneous dynamics observed in experiments is a consequence of the shape of the cell.

In this letter, we analyze the motion of two identical, self-avoiding chains, confined in a $2d$ rectangular box. Combining scaling theory, stochastic models, and Monte Carlo simulations, we show the existence of a transition from segregation to mixing with change of aspect ratio. The dynamics of segregation is shown to be inhomogeneous, with two distinct regimes. The segregation time depends sensitively on the geometry, and interestingly, exhibits a minimum at a specific aspect ratio that depends on the density.

\begin{figure}[htbp]
\centering
\includegraphics[width=0.85\linewidth]{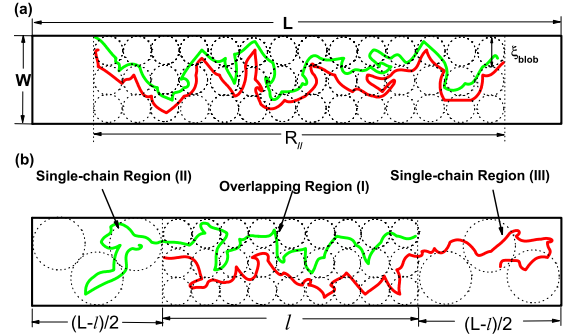}
\caption{(Color online)$(a)$ Blob structure of two polymers confined to a rectangular box. $R_{\parallel}$ is the extension of confined polymer (cf. text). The illustration is schematic and only the size of the blob, $\xi_{blob}$, is relevant. $(b)$ Illustration of box partitioning during segregation. Region I is overlapping and region II, III are single-chain. }
\label{fig:Confinement}
\end{figure}

Two identical self-avoiding chains with $N$ monomers are confined to a $2d$ box of width $W$ and length $L$, as illustrated in Fig.\ref{fig:Confinement}(a). The width is chosen to be less than $R_g$, the radius of gyration of  the free polymer.
A scaling theory of the free energy of confined polymers is based on the blob pictures with each blob contributing $k_BT$ to the free energy of a polymer (Fig.\ref{fig:Confinement}(a)) \cite{degennes-scaling, Sakaue-Flow, Suckjoon-ConfineSpace}. The blob size, $\xi_{blob}$, is determined by the competition between confinement, excluded-volume effects, and entropy. For length scales small compared to $\xi_{blob}$, confinement effects are screened and self-avoidance effects dominate, but for length scales larger than the blob size, polymer conformations are strongly affected by the confining geometry.

For a single chain confined in a rectangular box, assuming homogeneous distribution of monomers \cite{degennes-scaling, Sakaue-Flow, Suckjoon-ConfineSpace}, leads to
\begin{eqnarray}
\xi_{blob} \propto (\frac{WL}{N})^\frac{\nu}{2\nu - 1} ~,
\label{xibulk}
\end{eqnarray}
where  $\nu = \frac{3}{4}$ is the Flory exponent in $2D$, and the length is measured in units of the Kuhn length \cite{Sakaue-Flow, degennes-scaling}. The width of the box provides an upper bound for $\xi_{blob}$, which defines a length scale $R_{\parallel} \propto NW^{-\frac{1}{3}}$ \cite{degennes-scaling} that measures the extension of the polymer in the longitudinal direction. In a rectangular box, if $L >> R_{\parallel}$, blobs occupy only a small portion of the confining space, and the situation is similar to an open slit. In the opposite limit, two-dimensional confinement should significantly affect the properties of the chain. We introduce a dimensionless parameter $\lambda \equiv \frac{2R_{\parallel}-L}{L}$, to measure the influence of the closed geometry; $\lambda \geq 1$ if $L \leq R_{\parallel}$, and $\lambda \leq 0$ if $L \geq 2R_{\parallel}$.
Since each blob contributes $\sim k_BT \equiv \frac{1}{\beta}$ to the free energy of a single polymer \cite{degennes-scaling, Sakaue-Flow}, $F(L,W,N)$:
\begin{eqnarray}
\beta F(L,W,N) \propto N_{blob} =
\begin{cases}
\!\!\frac{N^3}{(LW)^2} &\mbox{if } \lambda \geq 1\\
\!\!\frac{N^3}{(R_{\parallel}W)^2} = NW^{-\frac{4}{3}} &\mbox{if }  \lambda \leq 1
\end{cases}
\label{GeneralFreeEnergy}
\end{eqnarray}
where $N_{blob}$ is the number of blobs in a single confined polymer.

In order to model two overlapping polymers, as illustrated in Fig.\ref{fig:Confinement}(a), the blob picture can be extended, assuming (a) that the density is uniform, and (b) that $R_{\parallel}$ is the same as that of a single polymer (limit of infinitesimal thickness of chains). The monomer linear density for two overlapping polymers is:
\begin{eqnarray}
\phi =
\begin{cases}
\frac{2N}{L}     &\mbox{if } \lambda \geq 1\\
\frac{2N}{R_{\parallel}} &\mbox{if } \lambda \leq 1
\end{cases}
\label{MonomerDensity}
\end{eqnarray}
During the process of segregation, the box space is naturally partitioned into three regions: Region I of length $l$ where the two polymers overlap, and Regions II and III that are occupied by single-polymer segments (Fig.\ref{fig:Confinement}(b)). By symmetry, regions II and III are equivalent.  The total free energy of the polymers is a function of the overlap distance $l$, $0 \leq l \leq R_{\parallel}$; $F(l) = F_{o} + 2F_{s}$ where $F_{o}$, $F_{s}$ denote the free energy in the overlapping and single-chain regions, respectively.  If the dominant mechanism of segregation is a lateral sliding of the polymers with negligible transverse displacements, then the linear density in Region I remains fixed at $\phi$, defining the monomer densities $\phi_{I} = \phi$, $\phi_{II} = \phi_{III} = \frac{2N - \phi l}{L - l}$. The different densities in regions I and II(III) imply different blob sizes (Fig. \ref{fig:Confinement}(b)).  Using Eq. \ref{GeneralFreeEnergy}, we obtain $F_o = F(l,W,\phi l)$, $F_s = F(\frac{L-l}{2},W,\frac{2N-\phi l}{2})$.
For $\lambda \geq 1 $,
\begin{eqnarray}
\beta F(l) \propto \frac{8N^3}{L^2W^2} ~,
\label{FreeEnergyStrong}
\end{eqnarray}
and for $0 \leq \lambda \leq 1 $
\begin{eqnarray}
\beta F(l) \propto
\begin{cases}
\frac{2}{W} (R_{\parallel}+3l) &\mbox{if } 2R_{\parallel}-L \leq l \leq R_{\parallel}
\\
\frac{8}{W}[\frac{(R_{\parallel}-l)^3}{(L-l)^2}+l] &\mbox{if } 0 \leq l \leq 2R_{\parallel}-L
\end{cases}
\label{FreeEnergyWeak}
\end{eqnarray}
For $\lambda \leq 0$, $\beta F(l) \propto \frac{6l}{W}$ is identical to the free energy of two chains in an open tube \cite{Suckjoon-segregationopen},   and the longitudinal confinement has no effect.  For $\lambda \geq 1$, the total free energy is independent of $l$, therefore, there is no force driving segregation and the motion of the two chains is purely diffusive \cite{Suckjoon-ConfineSpace}. For $\lambda \leq 1$, $F(l)$ is a monotonically increasing function of $l$,  and this repulsive potential drives segregation \cite{footnote}.

The blob-structure-based free energy function (Eq. \ref{FreeEnergyWeak}), can be used to model the long-time behavior of the segregating polymers.  The segregation time $\tau$ is defined as the average time for chains moving from $l = R_{\parallel}$ to $l = 0$, which is equivalent to the mean first-passage time of the corresponding Fokker-Planck equation \cite{Risken-FokkerPlanck, Grosberg-FirstPassage, Sung-Park}.
\begin{eqnarray}
\frac{\partial P(l,t)}{\partial t} = \frac{\partial}{\partial l}D e^{-\beta F(l)}\frac{\partial}{\partial l}  e^{\beta F(l)}  P(l,t) ~,
\end{eqnarray}
where $P(l,t)$ is the probability of the two chains overlapping by $l$ at time $t$,  and $D$ the diffusion constant given by the Einstein relation, $D = \frac{k_BT}{N\zeta}$ with $\zeta$ being the monomer friction coefficient \cite{Sung-Park, Kardar-AnomalousTranslocation}. This description is valid if  the segregation process is sufficiently slow such  that at each stage, the motion of segments is controlled by equilibrium statistics \cite{Risken-FokkerPlanck, Sung-Park, Kardar-AnomalousTranslocation}.
The global relaxation time of the end-to-end distance of a confined polymer grows less rapidly than  $N^2$ \cite{Arneodo-relaxation,Walter-Keith}. As shown below, the segregation time in the regime $\lambda \simeq 1$, grows faster than $N^2$, making the equilibration assumption reasonable.  Since $F(l)$ is a piecewise continuous function with different expressions in Regions I and II(III), the mean first-passage time can be written as the sum of  the first-passage times of two subprocesses: $\tau = \tau_{1} + \tau_{2}$, where $\tau_1$ is for the subprocess of separation between $l=R_{\parallel}$ and $l=2R_{\parallel} -L$ due to the entropy, and $\tau_2$ is for separation between $l=2R_{\parallel} -L$ and  $l=0$\ due to the competition between the entropic force and longitudinal pressure \cite{Kolomeisky-translocation, Muthukumar-translocation}.  Introducing $x \equiv R_{\parallel} - l$, we obtain, modulo constant factors:
\begin{eqnarray}
\tau_{1} &=& \frac{1}{D}\int_0^{L - R_{\parallel}}e^{\beta \mathcal{F}_{1}(x)} \int_0^x e^{-\beta \mathcal{F}_{1}(y)}\,dy\,dx \nonumber \\
\tau_{2} &=&
 \frac{1}{D}\int_{L - R_{\parallel}}^{R_{\parallel}}e^{\beta \mathcal{F}_{2}(x)} \int_{L - R_{\parallel}}^x e^{-\beta \mathcal{F}_{2}(y)}\,dy\,dx
\label{Eq. SegregationTime}
\end{eqnarray}
where $\beta\mathcal{F}_{1}(x) = -\frac{6}{W}x$, $\beta\mathcal{F}_{2}(x) = \frac{8}{W}(\frac{x^3}{(L - R_{\parallel} + x)^2} - x)$ \cite{Sung-Park,Muthukumar-translocation}.

In the limit of $\lambda \approx 0$ ($L \approx 2R_{\parallel}$), the segregation time is governed by $\tau_{1}$:
\begin{eqnarray}
\tau_{1} &=&   \frac{W^2}{36D}[e^{-\frac{6(L-R_{\parallel})}{W}} + \frac{6(L-R_{\parallel})}{W} - 1] \nonumber \\
 &\approx& \beta \zeta WN(L-R_{\parallel})
\label{Eq.Tau1}
\end{eqnarray}
In the second equation, the limit $\frac{L-R_{\parallel}}{W}\approx \frac{R_{\parallel}}{W} >> 1$ is taken.
Comparison with the time scale for pure diffusion from overlapping to separation $\tau_{diff} \sim \beta \zeta NR_{\parallel}^2$ \cite{degennes-scaling},  shows that the segregation time is much smaller than $\tau_{diff}$.
In the limit of $\lambda \approx 1$, $\tau_{2}$ is dominant. Since $\beta\mathcal{F}_{2}(x)$ is a gradually decreasing function with changes of the order of $ (1- \lambda)^2$, a linear approximation can be used to estimate $\beta\mathcal{F}_{2}(x)$:
\begin{eqnarray}
\tau \approx \frac{\beta \zeta W^{\frac{1}{3}}N^3}{(L-R_{\parallel})}
\label{Eq.Tau2}
\end{eqnarray}
Blob scaling arguments for free energy are only  valid for $(L-R_{\parallel}) > W$,  which provides an upper limit: $\tau \leq \frac{\beta \zeta W^{\frac{1}{3}}N^3}{W} \simeq \tau_{diff}$. Therefore, as $L \rightarrow R_{\parallel}^+$, $\tau \rightarrow \tau_{diff}$, the segregation dynamics approaches the unbiased-diffusion limit.   Since $\tau_1$, $\tau_2$ are decreasing and increasing functions of $L-R_{\parallel}$, respectively, the segregation time can exhibit a minimum as a function of $L$ for a fixed $N$ and $W$.

These theoretical predictions, were tested using  Monte Carlo simulations based on the bond fluctuation model (BFM) \cite{Binder-BFM, Kremer-BFM}. The BFM is a coarse-grained model of polymers in which chains live on a hypercubic lattice and fluctuations on scales smaller than the lattice constant are suppressed. The polymer is represented by a chain of effective monomers connected by bonds which are constructed to account for excluded-volume effects. An overlapping configuration of two chains is created by introducing a pseudo harmonic interaction: $\sum_{i=1}^{N} k(\bold{R}_{1i}-\bold{R}_{2i})^2 $, where $\bold{R}_{mn}$ denotes the position vector of the $n$th monomer on the $m$th polymer and $k$ is a parameter controlling the attractive strength. This interaction is turned off after the two chains are fully relaxed in the overlapped configuration. The separation between two chains is measured  by the horizontal  ($X_{cc}$) and vertical ($Y_{cc}$) projection of the centers of mass, and $X_{cc}$ is related to $l$:
\begin{eqnarray}
X_{cc} =
\begin{cases}
\frac{1}{2}(L-l) &\mbox{if } 0 < l < 2R_{\parallel} - L\\
R_{\parallel}-l  &\mbox{if } 2R_{\parallel} - L < l < R_{\parallel}
\end{cases}
\end{eqnarray}
\begin{figure}
\centering
\subfigure{
\includegraphics[width=0.9\linewidth]{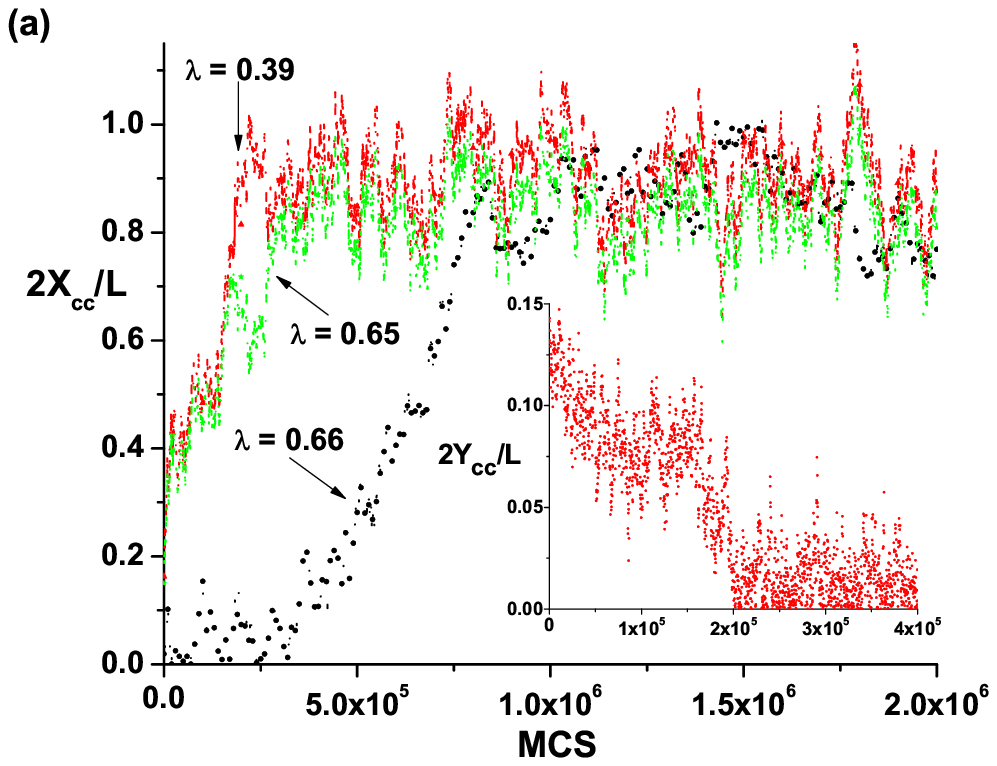}
}
\subfigure{
\includegraphics[width=0.9\linewidth]{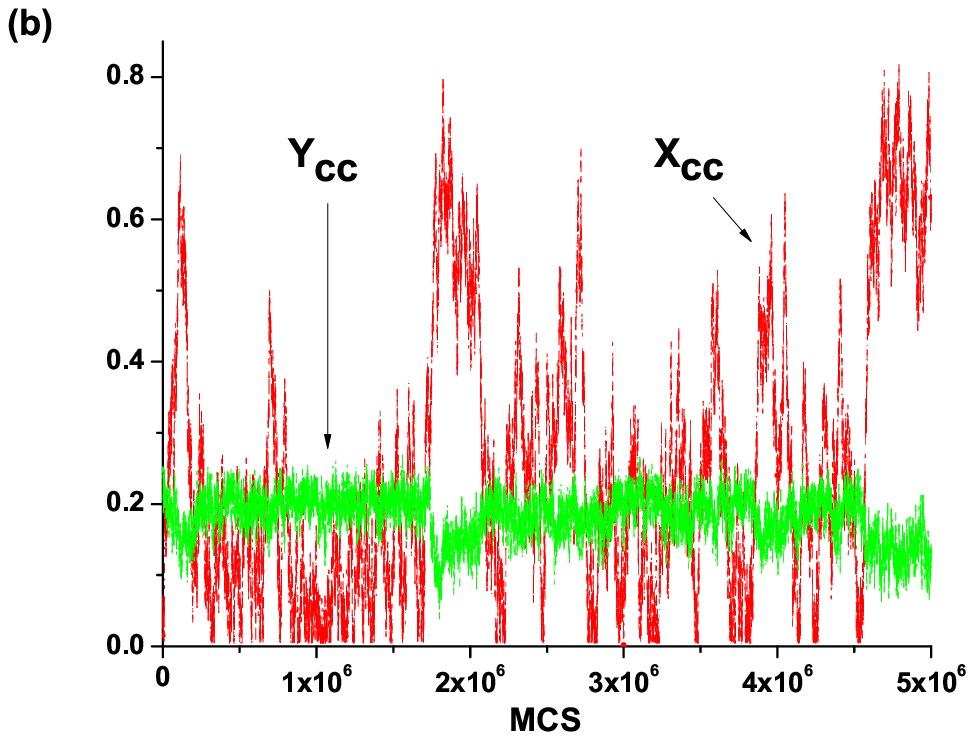}
}
\caption{(Color online)(a). Plot of $2X_{cc}/L$ vs. MCS in the segregation region for ($\lambda = 0.39$; $N = 100, L = 60$), ($\lambda = 0.65$; $N = 120, L = 65$), and ($\lambda = 0.66$; $N = 200, L = 100$). The inset shows $Y_{cc}$ vs. MCS. (b). Plot of $2X_{cc}/L$ and $2Y_{cc}/L$ vs. MCS in the mixing region: ($\lambda = 1.08$; $N = 100, L = 40$).}
\label{fig:Centerdistance}
\end{figure}

Simulations were performed over a wide range of parameters: $80 <N < 200$, $30 < L <140$,  keeping  $W=10$ fixed. Each Monte Carlo trajectory spans a few hundred Rouse times \cite{degennes-scaling, Binder-BFM}, and twenty independent trajectories are sampled for each set of parameters. We have checked the validity of the assumption in Eq. \ref{MonomerDensity} and find that the difference of average extension between single chain and two chains is less than 6\% \cite{Ya&Bulbul-Unpublished}. Fig.\ref{fig:Centerdistance} illustrates the evolution of $X_{cc}$ and $Y_{cc}$ with respect to Monte Carlo Steps (MCS).

As shown in Fig.\ref{fig:Centerdistance}(a), for  $\lambda = 0.39, 0.65, 0.66$, $X_{cc}$ grows and fluctuates around  $L/2$ and $Y_{cc}$ decreases to zero,  a signature of segregation. The time scale for reaching a well-defined average is the same for $X_{cc}$ and $Y_{cc}$, which is a convincing argument for equating this measured time to the calculated segregation time, $\tau$.  Fig.\ref{fig:Centerdistance}(b) shows that for $\lambda = 1.08$, $X_{cc}$ and $Y_{cc}$ do not grow but fluctuate between  $0$ and $\frac{L}{2}$ ($X_{cc}$), and between 0 and $\frac{W}{2}$($Y_{cc}$), indicating a lack of segregation.

\begin{figure}[htbp]
\centering
\includegraphics[width=0.95\linewidth]{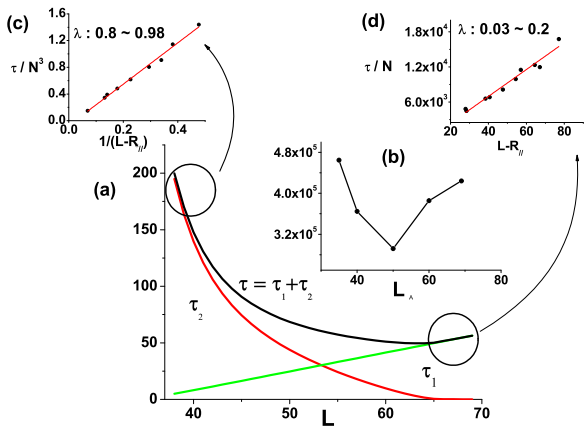}
\caption{(Color online)(a) Plot of $\tau_1$, $\tau_2$ and $\tau$ vs. $L$ obtained from numerical integration of Eq. \ref{Eq. SegregationTime} with $N = 80, W=10$. (b) Monte Carlo simulation results for the same set of parameters as (a). (c) Plot of $\tau/N^3$ vs. $1/(L - R_{\parallel})$ for $0.8 < \lambda < 0.98$. (d) Plot of $\tau/N$ vs. $(L - R_{\parallel})$ for $0.03 < \lambda < 0.2 $.  In (c) and (d), the lines denote two fits for the asymptotic theoretical predictions (Eq. \ref{Eq.Tau1}, \ref{Eq.Tau2}).}
\label{fig:RescaleTime}
\end{figure}
In Fig. \ref{fig:RescaleTime}(a),(b), Monte Carlo simulation results and numerical integrals of Eq. \ref{Eq. SegregationTime} are shown; both demonstrate non-monotonic behavior of $\tau$ as a function of $L$.  In Fig. \ref{fig:RescaleTime}(c),(d),  the segregation time $\tau$ is scaled according to the relations in Eq. \ref{Eq.Tau1} and \ref{Eq.Tau2}, the linear fits support the  theoretical predictions in the asymptotic regimes.

\begin{figure}[htbp]
\centering
\includegraphics[width=0.9\linewidth]{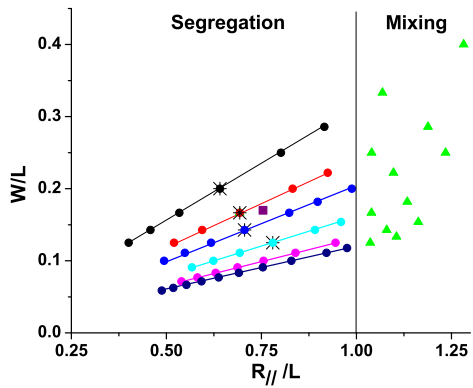}
\caption{(Color online) Phase diagram of segregation and mixing. The x-axis parameterizes the monomer concentration and the y-axis parameterizes the geometry. Simulation data are denoted by dots (segregation) and triangles (mixing). Lines illustrate data from various $L$: $N = 80$ (black), 100 (red), 120 (blue), 150 (cyan), 180 (magenta), and 200 (navy) from top to bottom. Stars mark aspect ratios with the minimal segregation times. The purple square denotes an estimate based on experimental data for $\textit{E.coli}$. }
\label{fig:Phasediagram}
\end{figure}

Fig. \ref{fig:Phasediagram} shows simulation data and the theoretical predicted phase boundary separating the segregated and the mixed phase, demonstrating the accuracy of the theoretical predictions. Along each line, which corresponds to a given chain length, the geometry with minimal segregation time is marked (Fig. \ref{fig:Phasediagram}). A minimal extension of the blob picture to $3d$ ring polymers such as the chromosome of $\textit{E.coli}$ is to require $\xi_{blob} \leq \frac{W}{2}$ rather than $W$ \cite{Ya&Bulbul-Unpublished}. Applying this model to $\textit{E.coli}$, using measured parameters \cite{JunRMP, Jun2006, Odijk-Genome}: $\xi_{blob} \approx 87nm, W \approx 0.24 nm$, and $L\approx1.39 \mu m$, locates $\textit{E.coli}$ in the segregation phase and close to the geometrical condition of minimum segregation-time. While this observation may be fortuitous since chromosome strand is immensely more complicated than a linear polymer, it raises the interesting possibility that genome segregation times could have applied evolutionary selection pressure to genome lengths.

In conclusion, we have investigated the dynamics of segregation of two self-avoiding chains  confined to  a narrow rectangular box. A theoretical framework, based on the blob picture to capture the essence of the competition between excluded volume and confining effects, predicts  a rich phenomenology of transitions between segregated and mixed states and optimal geometries that minimize the segregation time.   Monte Carlo simulations provide broad support for the theoretical predictions.
Experiments  in microfluidic devices should be able to provide direct tests of the predictions, and elucidate the role of entropy  in driving segregation of biopolymers. The simulation can be extended to study more realistic models based on actual chromosomes structure of bacterial.

This work was supported by the Brandeis MRSEC (NSF-0820492). The authors acknowledge helpful discussions with Suckjoon Jun, Mike Hagan and Jan$\acute{e}$ Kondev.

\bibliographystyle{apsrev}
\bibliography{Segregation-bc}

\end{document}